# Processing Color in Astronomical Imagery

Kimberly K. Arcand[1], Megan Watzke[1], Travis Rector[2], Zoltan G. Levay[3], Joseph DePasquale[1], & Olivia Smarr[4]

[1]Smithsonian Astrophysical Observatory, 60 Garden St, Cambridge, MA 02138, USA

[2]University of Alaska Anchorage, 3211 Providence Dr. Anchorage, AK 99508, USA

[3]Space Telescope Science Institute, 3700 San Martin Drive, Baltimore, Maryland 21218, USA

[4]Stanford University, 531 Lasuen Mall, Stanford, CA 94309, USA

Correspondence: Kimberly K. Arcand, Smithsonian Astrophysical Observatory, 60 Garden St, Cambridge, MA 02138, USA. Tel: 1-617-218-7196. E-mail: kkowal@cfa.harvard.edu

**Abstract**

Every year, hundreds of images from telescopes on the ground and in space are released to the public, making their way into popular culture through everything from computer screens to postage stamps.

Most astronomical images are natively digital, with their data traveling from telescopes to scientists as a series of ones and zeroes and with tens and even hundreds of terabytes of data available from a single observatory archive (Brunner, 2001). These data span the entire electromagnetic spectrum from radio waves to infrared light to X-rays and gamma rays – a majority of which is undetectable to the human eye without technology. Once these data are collected, one or more specialists (scientists and other professionals) must process the data to create an image. Therefore, the creation of astronomical imagery involves a series of choices. How do these choices affect the comprehension of the science behind the images? What is the best way to represent data to a non-expert? Should these choices be based on aesthetics, scientific veracity, or is it possible to satisfy both?

This paper reviews just one choice out of the many made by astronomical image processors: color. The choice of color is one of the most fundamental when creating an image taken with modern telescopes. We briefly explore the concept of the image as translation, particularly in the case of astronomical images from invisible portions of the electromagnetic spectrum. After placing modern astronomical imagery and photography in general in the context of its historical beginnings, we review the standards (or lack thereof) in making the basic choice of color. We discuss the possible implications for selecting one color palette over another in the context of the appropriateness of using these images as science communication products with a specific focus on how the non-expert perceives these images and how that affects their trust in science. Finally, we share new data sets that begin to look at these issues in scholarly research and discuss the need for a more robust examination of this and other related topics in the future to better understand the implications for science communications.

## 1. Image as Translation

Astronomy is largely funded by taxpayer dollars, and because the nature of the scientific process includes the dissemination of results—in the form of proof or a "public report" (Miller, 2004: 279)—many in the professional community believe in the responsibility of releasing meaningful results to the public (Frankel, 2004: 418). Quite often, the images produced from astronomical data are a great asset in doing this; they provide a principal source of information that forms public conceptions about space (Snider, 2011). One can draw a parallel from Felice Frankel's work in creating visually appealing images of materials science and the importance of producing "wonderful and accessible images" to effectively communicate science (Frankel, 2004: 418).

With advancements in detector technology, many astronomical images created today sample data from wavelengths outside those of human vision. These data must be translated, through the use of processing, adding color and artifact removal, smoothing and/or cropping into human-viewable images. The choices involved in deciding how best to represent or translate such data underscore the idea that "what we encounter in the media is mediated" (Mellor, 2009: 205).

There is, however, a lack of robust studies to understand how people—particularly non-experts—perceive these images and the information they attempt to convey. The extent to which the choices—including those surrounding color—made during image creation affect viewer perception, comprehension, and trust is only beginning to be more closely examined. For example, the influence of images in news reports has not been extensively studied (Jarman et al., 2011: 4). Are there rules when it comes to presenting something visually that cannot be seen with the human eye? Is there an inherent amount of manipulation through the choice of color in this process of creating astronomical images? Or is it simply providing a proxy so that we can experience something (that is, views of the universe) that otherwise may not be available to us?

The intent behind the manipulation of astronomical data for the creation of public images is important. The choices made by the scientist must not, of course, change the overall interpretation (Rossner and Yamada, 2004: 11). The assumption made here is that the intent is to add value and information, make the presentation of the data aesthetically pleasing, make the information accessible, and/or make the material understandable. Images can, after all, be both informational and beautiful (Frankel, 2004: 418). Color plays a vital role in these goals.

But to manipulate an image responsibly, the scientist or science image processor must also understand the perspectives of non-experts and be willing to embrace openness and transparency in their work to foster trust with the public (Irwin, 2009: 7). We argue that non-experts may not necessarily perceive colors in the same way that the scientific community does. In fact, until recently, there has been little scholarly research into the differences in perception of color between lay audiences compared with professional scientists, in particular, astronomers and astrophysicists. In this paper, we describe the current state of affairs in the creation of astronomical images using color, raise some of the issues in doing so, and outline some of our efforts to address these questions using research-based methodologies.

*1.1 Historical Notes, and Image as Truth*

Photographs, though made by human hands using machinery humans have invented, have been perceived as being free of human bias (Schwartz, 2003: 28). Photographs have also been considered as "records of the real" (Rothstein, 2010). Photography's veracity, though once seen as an inherently true vehicle of a snapshot in time, has seemingly decreased since the purported "reintroduction of the human hand" (Schwartz, 2003: 30) through the ability to manipulate images quickly and efficiently with digital imaging software such as Adobe Photoshop. There is, understandably, a negative connotation to manipulating photos, so much so that "to photoshop" has become a pejorative verb.

Since image manipulation is more convenient and accessible today, it is perhaps more in the public consciousness than it was with analog (film) processing. However, such control over an image's appearance has been around since the beginnings of photography. Aside from deceptive or entertaining uses of altering photographs, a skilled darkroom technician can evoke detail and tonality from the negative that is not apparent in a "straight" print, using techniques including altering exposure and development (contrast), dodging, burning, intensification, toning, etc. Ansel Adams, for example, was well known for his ability to adjust his processing from exposure through printing, even though many non-experts would likely consider his photos to be faithful representations of well-known landscapes (Adams and Baker, 1995).

The correspondence between photographs of scenes and how the eye sees those same scenes still affirms the general "truth" of photographic images today. The question of "is this what it really looks like" is commonly heard from the non-experts who look at astronomical images[i]. How does the color mapping being used to create those astronomical images play into this assumed truth? To explore this further, one needs to understand the origins of the astronomical image from the first telescopes.

## 2. Creation of the Astronomical Image

The history of astronomical imaging arguably began in 1609 when Italian astronomer Galileo Galilei, with his new astronomical telescope[ii], observed craters on the surface of Earth's Moon (figure 1a) and discovered four lunar bodies orbiting Jupiter. He then transcribed them into drawn format and published his findings in 1610[iii].

Astronomical imaging has been evolving ever since with the development of increasingly sophisticated technology. For example, during the 20th century, astronomers began making three-color composite images using filters on film. (Filters in this case refer to ways to limit the amount of light in each image. For example, some filters enable the astronomers look at the light emitted by only hydrogen atoms, or in other instances, in the higher energies detected by the telescopes.) Making three-color astronomical images involves stacking three different images of the same field of view of different slices of light, typically assigning red, green, and blue as the colors for each layer.

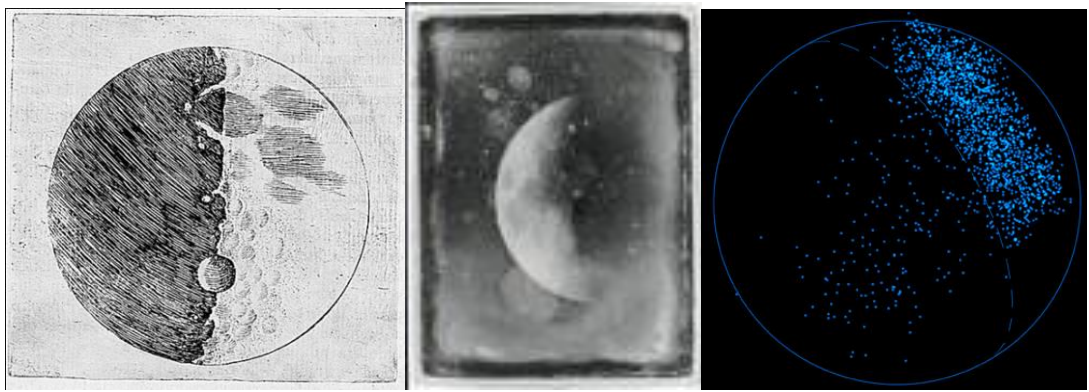

Figure 1a (left). Galileo's transcription of our Moon's surface from *Sidereus Nuncius* in the 17[th] century. Image courtesy of the Master and Fellows of Trinity College, Cambridge. Figure 1b (middle). One of the oldest surviving daguerreotype images of the Moon, from 1852, taken by John Adams Whipple (Credit: Harvard/Wikimedia Commons http://commons.wikimedia.org/wiki/File:1852_Moon_byJAWhipple_Harvard.png). Figure 1c (right). The Moon in X-ray light, taken in 2001 by the Chandra X-ray Observatory. (Credit: NASA/CXC/SAO)

During the second half of the 20[th] century, there were major advances for astronomical telescopes on the ground with larger and more sophisticated mirrors being constructed (DeVorkin, 1993). This period also saw the dawn of the Space Age (Portree, 1998), of which astronomers were eager to take advantage (for example, Tucker and Tucker, 2001: 28). By being able to launch telescopes into space, astronomers could build detectors and other instruments that were sensitive to ultraviolet, X-ray, and other types of light that are absorbed by the Earth's atmosphere. Given these new technologies, and specifically since the launch of NASA's Hubble Space Telescope, the period since the 1990s is considered a "new era" for astronomy (Benson, 2009: 322). There are now many space-based telescopes and observatories on the ground available to look at a wide range of different types of light from the cosmos. Astronomical data from such modern telescopes are no longer captured on film. They are instead natively digital and arrive from the telescope in a form that generally requires processing in order to create an image.

This boon in astronomical data also means that today we find popular media flush with astronomical images. But the transition in science image communication from hand-drawn observation to emerging photomechanical techniques in astrophotography and the most advanced space-based optics today reflects a shift in the purported subjectivity of the scientist when producing these images as vehicles of visualizing his or her data. Mechanically produced images were assumed to eliminate the "meddling" mediation of the human hand—an instrument having no will, desires, or morals of its own (Daston and Galison, 1992: 83). But as the next section notes, complete objectivity and neutrality in the interpretation of scientific data into the pictorial form is not feasible, nor is it desired. The science embedded in the image often informs the human choices for the color composite.

*2.1 The Processing Pipeline*

2.1.1 Translation from 1's and 0's into Pixels

As mentioned previously, modern astronomical images—whether from an optical telescope like the Hubble Space Telescope or one that looks at X-rays like the Chandra X-ray Observatory—are inherently digital. When a satellite, for example, observes an object in space, its camera records the exposure of light-sensitive electronics to photons. These exposure recordings then come down to Earth from the spacecraft via a network in the form of 1's and 0's. Scientific software then translates that data into an event table that contains the position and, in some cases, certain information for the time and energy, of each photon that struck the detector during the observation. The data is further processed with software to form the visual representation of the object as pixels (Rector, 2007: 599).

2.1.2 Mapping Color

Color plays an important role in human perception and culture, and also has some meaning in the realm of science. The spectrum of light that can be detected with the human eye can be broken into various colors from red to green to violet. In astronomy, these colors are applied to wavelengths of light far beyond the range humans can see. Using telescopes in space and those on the ground, astronomers can now "see" from radio waves to gamma rays, across the full range of the electromagnetic spectrum (see figure 2).

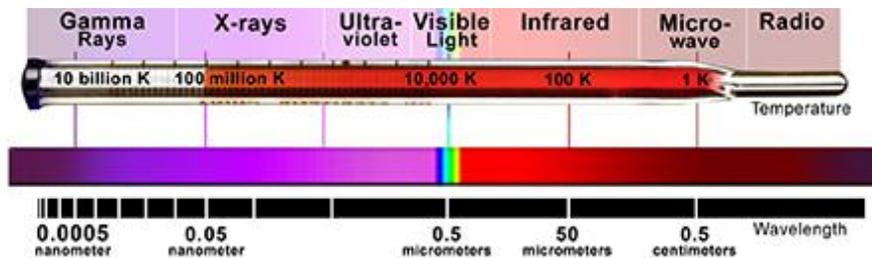

Figure 2. Electromagnetic radiation comes in a range of energies, known as the electromagnetic spectrum. The spectrum consists of radiation such as gamma rays, X-rays, ultraviolet, visible, infrared, and radio.
(Illustration: NASA/CXC/SAO)

The data that are received by telescopes originates as variations of brightness with no inherent color. Color must then be added to the image. To do this, scientists or science image processors put together a color map that "translates" portions of the data to various colors that can be seen with the human eye.

Here we take this opportunity to raise the issue of false color. The term false color is often used to describe astronomy images whose colors represent measured intensities outside the visible portion of the electromagnetic spectrum[iv]. But a false color image is not wrong or phony as the term might imply (Hurt, 2010). It is a selection of colors chosen to represent a characteristic of the image (intensity, energy, or chemical composition, for example). False color images are not actually "false in any sense of the word, but only a matter of different visual codes" (Snider, 2011: 9). The colors used are representative of the physical processes underlying the objects in the image[v]. Alternatively, a scientist or image processor may want to be able to differentiate between low, medium, and high energy, assigning red, green, and blue (RGB) in a "chromatic order" (Wyatt, 2010: 35) from low to high energy as it does in a rainbow.

This use of RGB for low, medium, and high energies can be applied to wavelengths that fall outside human vision. For this first example, we look at the black hole at the center of the Milky Way, known as Sagittarius A* (figure 3a) as observed by the Chandra X-ray Observatory. In this image, the X-rays near the black hole (seen as the bright white structure in the center) and greater surrounding area are shown in three colors: red (2-3.3 keV), green (3.3-4.7 keV), and blue (4.7-8 keV)[vi]. All of these X-ray wavelengths are invisible to the human eye but can be seen in relation to one another through an application of color.

As we mentioned earlier, some astronomical images are colored to highlight certain scientific features in the data. For example, an image of supernova remnant G292 (figure 3b) shows colors that were selected to highlight the emission from particular elements such as oxygen (yellow and orange), neon (red), magnesium (green), and silicon and sulfur (blue)[vii]. Again, this is an image made from X-ray light. Each color represents the specific wavelengths of light that each element gives off or is absorbed in this environment. Compare figures 3b and 3c; the colors added in 3b provide additional insight into physical characteristics of the X-ray gas in this supernova remnant that could not be gleaned from the black and white image.

In both cases, the choice of color is contributing to the informational quotient of the image because the colors reflect the processes inherent in these objects. When color maps are combined, they yield "new insights into the nature of the objects" (Villard, 2010). However, since the selection of wavelengths and decisions on color are not standardized, it may be confusing to non-experts and lead to perceptions of phoniness in the images.

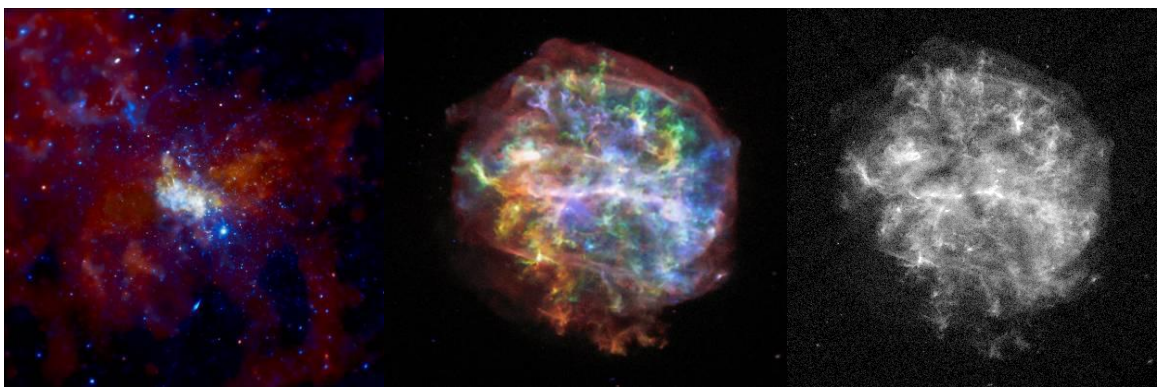

Figure 3a (left) Sagittarius A* in X-ray light. Figure 3b (middle) Supernova remnant G292 in X-ray light. Figure 3c (right) Raw, unprocessed broadband X-ray image of supernova remnant G292.
(Credit: NASA/CXC/SAO)

How does all of this relate to the ethical issues of color representation? For one, it exemplifies the idea that scientists and their collaborators make use of interpretation, selection, judgment, and artistry (Daston and Galison, 1992: 98) when deciding how to represent specific scientific data. To put it another way, both scientific and aesthetic choices are being made when an image is created.

Modern astronomical images function as "an illustration of the physical properties of interest rather than as a direct portrayal of reality as defined by human vision" (Rector, 2005: 197; see also Rector, 2007). The next section of this paper examines the various degrees of manipulation possible in the color mapping of astronomical images and whether those decisions might be perceived as ethical by non-experts.

## 3. Other Colorful Examples

### 3.1 A Recognizable Object Seen Differently

While perhaps not considered by the greater public to be an astronomical object, the Sun, of course, is our closest star. The most common view of our Sun is that as seen from the ground: an unchanging yellow disk. Images from NASA's Solar Dynamics Observatory (SDO) reveal the Sun (figure 4a) in a much different way. Using detectors that observe in 13 distinct wavelengths from X-ray to ultraviolet, the space-based SDO captures curving and erupting prominences, and the features in the images trace magnetic field structure. The hottest areas appear almost white while the darker areas indicate cooler temperatures on our local star. The colors assigned as presented in this tableau of solar images[viii] are somewhat arbitrary in hue and saturation. However, when colors are combined to create a multiwavelength view of the Sun, colors are chosen in an RGB ordering according to temperature. In the single wavelength views, these images seem to "flaunt" their color (Wyatt, 2010: 35), almost purposefully making the familiar unfamiliar.

Is this more acceptable because the Sun is an easily recognized object by most citizens of our planet? It could be that because the color is so unexpected, it immediately functions as a visual cue that the color is not being presented as reality but rather as a layer of information for the viewer. Similarly, this might be part of the reason why brightly false-colored radar maps on the nightly news are acceptable (figure 4b). Foreign colors on recognized objects have less of a claim to veracity. Their point is, perhaps, clearer to the viewer.

If one observes, however, that images "are not self-explanatory" but rather that they "need to be interpreted, and the human task of interpretation is often a [big] obstacle" (Gladwell, 2004: 2), then there is still a risk of non-experts misunderstanding the purpose of color, even on a recognizable object that is not being understood as "true". This is particularly the case when no explanation of the color table is offered with the images. The satellite imagery for weather forecasts still utilizes color for a specific purpose—indicating variations in temperature or precipitation, for example—and either it is accompanied with a color legend or with verbal or textual narration. Any solar images would also benefit from clear, descriptive information to explicitly relate the purpose of the color.

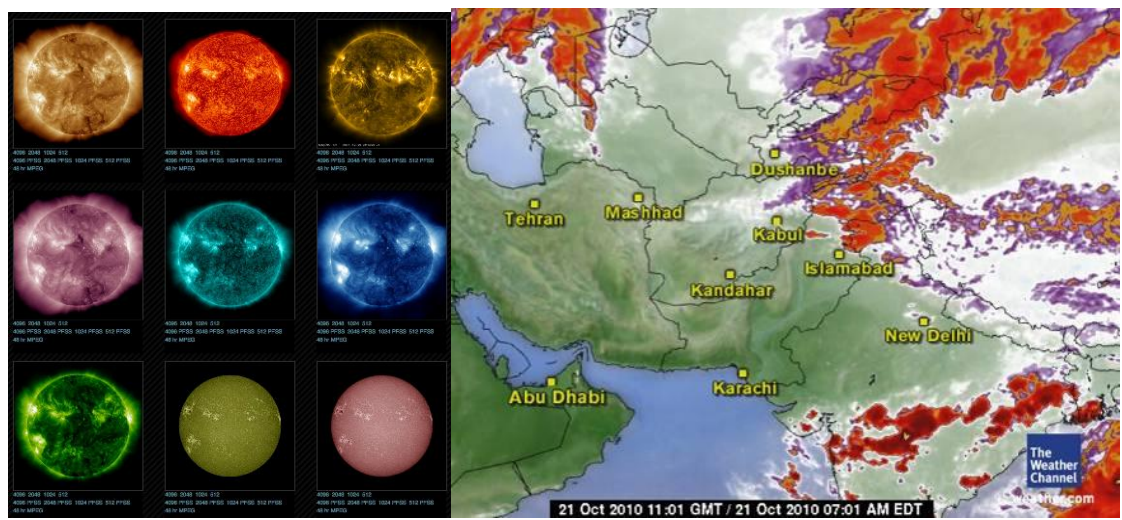

Figure 4a (left) Multi-bandpass images of the Sun. (Courtesy of NASA/SDO and the AIA, EVE, and HMI science teams.). Figure 4b (right) Satellite imagery of Asia. (Credit: The Weather Channel)

In another sense, the recognizable object can be something that grounds an unfamiliar and exotic process to a shared experience or common conception through the use of the visual metaphor (Frankel, 2002: 254). For example, figures 5a and 5b present a 2010 Chandra X-ray Observatory press release that shows a multiwavelength view of the massive galaxy M87. The galaxy harbors a super massive black hole at its center, producing massive jets of energetic particles, creating shockwaves that ripple throughout the galaxy. This process was compared with images and video of the Icelandic volcano Eyjafjallajokull, which erupted in 2010 and showed analogous shockwaves to those seen in M87. The color choice for the galaxy image added to the analogy with warm and cool hues providing contrast to the shockwaves in a similar way as was seen in the volcano imagery.

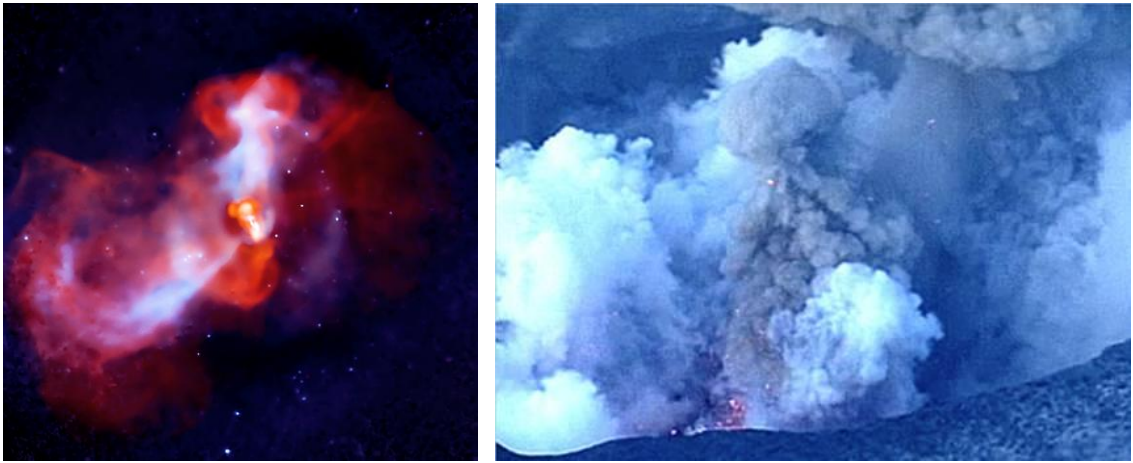

Figure 5a (left) M87 in X-ray (blue) and radio (red) light. Figure 5b (right) Icelandic volcano Eyjafjallajokull. (Credit: NASA/CXC/SAO; Volcano image: Omar Ragnarsson)

*3.2 Cultural Norms and Misconceptions*

NGC 4696 is a large elliptical galaxy in the Centaurus Galaxy Cluster, about 150 million light years from Earth. This composite image that was released to the public (figure 6a) shows a vast cloud of hot gas (X-ray = red) surrounding high-energy bubbles (radio = blue) on either side of the bright white area around the supermassive black hole[ix].

Astronomers and other experts in this area understand that the color blue physically represents higher temperatures than red does. (Blue stars, for example, can burn at temperatures that are tens of thousands of degrees higher than red stars.)

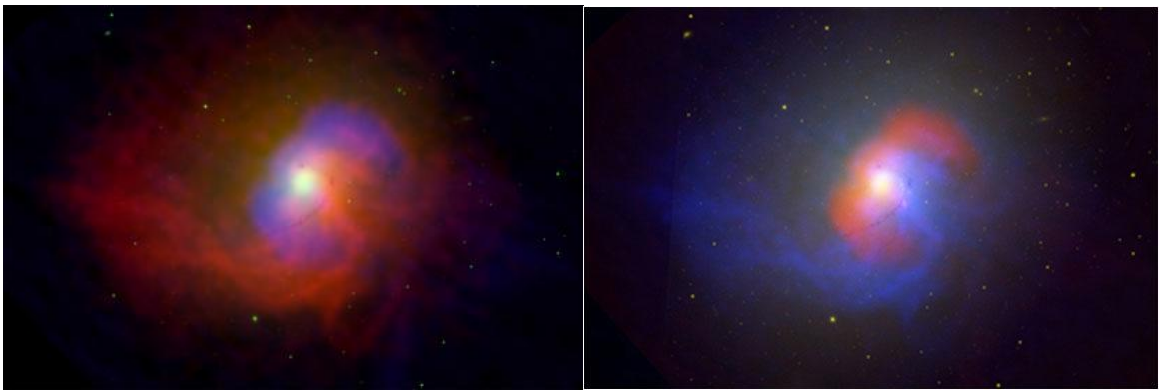

Figure 6a. Alternating colors: red (Figure 6a, left) and blue (Figure 6b, right) color-coded versions of galaxy NGC 4696 in X-ray, infrared, and radio light. (Credit: NASA/CXC/SAO)

However, most non-experts consider red to be hotter than blue. In a study done on this particular image of NGC 4696, 71.4% of the non-scientist study participants thought red was hotter than blue (Smith et al., 2010). To get across the point that this image is an immensely hot cloud of gas, the image processors inverted the more traditional, physically aligned color scheme and colored the highest-energy/hotter-temperature features red (figure 6a). Figure 6b shows the color mapping that is more typical of astronomical images, where the band passes are in chromatic order (red, green, and blue are mapped from low to medium to high energies, respectively).

The use of color here might be viewed as perpetuating a misconception. One could question if it is the responsibility of the scientist to correct or disabuse common notions, such as that red is hotter than blue. The primarily red image seems to convey—in a cultural language—the extreme heat of this object better, even though its color mapping might be considered by the scientific community to be non-standard. The responsibility may in fact fall to both scientists and the public to develop a mutually consistent visual literacy, a framework from which to view and understand these images (Trumbo, 1999).

Is it the case that one must "be true to the real nature of science, not to an idealized preconception" of that science (Ruse, 1996: 304)? Or can one also "acknowledge the realities of human nature" and make choices for the greater good—that is, choices that reflect the greater number of people who would then understand that image in a more meaningful way (Ruse, 1996: 304)? Aligning these images to cultural norms (regarding the perception of color, at least) might make the image and the science more trustworthy and relatable for the non-expert. Though it may seem counterintuitive not to explicitly follow the highest degree of "truth" represented in the data of a scientific image, the benefits of following cultural norms could outweigh the risks—particularly when the reasoning behind the choice is properly explained.

## 4. Discussion: Color and Construction

The examples above are just a small sample of the different ways astronomical data can be presented. The images are constructed from actual data. They look different due to the subjective choices of the scientist and science image processors—including choices of color. As Trumbo notes, "these seemingly simple considerations determine what the viewer encounters and ultimately contributes to what the viewer understands" (Trumbo, 2000: 387). How is it possible to maintain (or obtain) public trust in colored science images such as these?

Perhaps the way to do this is two-fold: through transparency and a more active provision of information. If non-experts can develop a visual literacy through access to the choices made—and to the information on why these choices need to be made in the first place—then it may be possible to decrease any feelings of the public being misled (at best) or deceived (at worst). Opportunities to provide viewers with a more complete narrative of the choices being made might include systematically posting raw (unmanipulated), black and white versions of the data that the scientist first analyzes. It could also be helpful to provide more opportunities for non-experts to see behind any *"Wizard of Oz"* curtain directly and be able to manipulate astronomy images themselves using free and open-source software, such as some observatories are starting to do[x].

Some steps are being taken to address these questions through analytical research. In 2008, the Aesthetics and Astronomy[xi] (A&A) project at the Smithsonian Astrophysical Observatory was formed to help better understand and clarify how non-experts perceive astronomical images. The goal of this project is to help provide scientists and science communicators with a more informed consensus on the perspectives of non-experts and help make these questions become less of an ethical issue. The A&A program consists of experts from a variety of fields including astrophysicists, education and public outreach professionals, and aesthetics experts from the field of psychology. By conducting both online and focus group-based studies, the A&A has begun to tackle the issues of color in astronomical images.

A project such as A&A represents an important step in better understanding how people of various backgrounds consume and relate to scientific images. The visual products of science are embedded with a sense of authority of the science itself. The color images are created to represent and visualize the science, but they are seldom actually used to do the science. There are rigorous techniques used by researchers to analyze their data. In other words, the scientific results are not subjective in the way the color images are. This authority is "enhanced by the traditional unidirectional model of science communication to the public" (Greenberg, 2004: 83). This is likely even more of the case when the process by which the image was produced is very complex—and with many choices having been made by the scientist—so much so that non-experts become less likely "to engage critically with it" (Greenberg, 2004: 83). That authority then, when unquestioned, can become more powerful. By not spelling out the various enhancements for the greater good, astronomy professionals risk embodying that lack of questioning as to how an image was made and how those decisions affected the final product. One of the goals of the scientist or science communicator is to help facilitate engagement with the audience in a "dialogic, 'two-way' or 'bottom-up' approach—sciences on one side, in a more symmetrical relationship with publics on the other—where openness and transparency are valued" (Holliman, Jensen, 2009: 37). One objective in creating images should be making the images and the science they embody accessible to the non-experts to which they are being presented.

## 5. Summary and Conclusion

Modern telescopes and the availability (online and otherwise) of the many resulting images have helped make astronomical imagery visible. People can experience a deep personal reaction to astronomical images (Arcand & Watzke, 2011), and some of that emotion may be influenced by the choice of color. Enhancing astronomical images with color adds to the information quotient of the image and enhances the aesthetic appeal. This aestheticism appeals to a pre-existing "sensibility" within the viewer and can elicit an emotional response that "entrances" or otherwise seduces the viewer (Hall, 1996: 28). These images are also of highly "exotic and inaccessible" (Hall, 1996: 20) processes and phenomena that make the pictorial representation of the data as well as visual metaphors important and useful. Taken in context with the text of a science article, these images "are not simply adjuncts to the written word; they are integral to the process of meaning making" (Jarman et al, 2011:4).

In order to help establish or maintain public trust in the representation of such scientific data, it is imperative that additional transparency is provided by the image creator to reveal details on how the images were made, including the choices of colors for each image and why those colors were chosen. If the greater public perceives an image to have been "faked" or otherwise believes the use of false color has no purpose but to mislead, then this undermines the credibility of the science that is being conveyed. Likewise, it is incumbent upon the public to develop and embrace a visual literacy to enable a critical reading of these images (Trumbo, 2000; Jarman et al, 2011). This issue of accountability and full disclosure is important to reconcile the perceived "truth" in an image derived from manipulated data.

Though color to the scientist might not be the most important factor of the research result, color to the non-expert is perceived as having importance. It is the way our minds function everyday in a cultural context, from the red traffic light telling a driver to stop to the yellow tinge of a baby's skin being a clue to jaundice or a green hue in the sky warning of bad weather. These colored pixels have power, and they convey meaning. These visual representations of data tell a story and say that the "role of the image in the communication of science is an increasingly complex one" (Trumbo, 2000:380).

In summation, for astronomy it is not feasible to have a rigidly standardized treatment of color due to the inherent complexity of the data and the many types of data being used. However, it should be possible to convey the imagery with such transparency that the aesthetic appeal is maximized while assuring the greater public that scientific integrity has been maintained, and subjectivity made plain, to aid in their interpretation and critical reading of the visual representation. Effectively communicating the science of the universe continues to be challenging, yet offers significant rewards for expert and non-expert alike.

## Acknowledgements

This material is partially based upon work supported by the National Aeronautics and Space Administration under contract NAS8-03060. Portions of this paper have been presented at Brown University (2010, 2011). Portions have also been used in an online exhibit (http://chandra.si.edu/art/color/colorspace.html) and article (http://www.huffingtonpost.com/megan-watzke/cosmic-microwave-background_b_3045048.html)     Special thanks to Dr. Sheila Bonde and Dr. Paul Firenze of Brown University, Dr. Peter Edmonds of SAO, and Dr. Lisa Smith of University of Otago, New Zealand for input.

http://news.discovery.com/space/astronomy/technicolor-universethis-week-astronomers-released-an-arresting-color-picture-of-a-pair-of-colliding-galaxies-combined-from.htm.

Wyatt, R. (2010). Visualising Astronomy: Let the Sun Shine In. *Communicating Astronomy with the Public, 8*, 34-35.

---

[i] For example, as noted during a focus group at the Smithsonian Astrophysical Observatory for the Aesthetics & Astronomy research project, December 2010.

[ii] Note that Thomas Harriot's telescope preceded Galileo's as the first astronomical telescope, though Galileo's findings are considered to be more significant. See Edgerton Jr, Samuel Y. 1984 (Fall), 'Galileo, Florentine "Disegno," and the "Strange Spottednesse" of the Moon', Art Journal, Vol. 44 (3) pp. 225-232

[iii] Galileo's drawings of the moon allowed him to argue that the lunar surface was filled with mountains and valleys.   This was in direct contradiction with Aristotelian doctrine stating that all celestial bodies were perfectly smooth.   Whitaker, E.A. 1978, *Journal for the History of Astronomy*, **9**, 155-169

[iv] See also http://chandra.si.edu/photo/openFITS/overview.html by Chandra EPO Group at SAO (DePasquale, Arcand, Watzke et al 2009)

[v] It might be desirable to replace the term false color with representative color to help remove some of the negativity that the inclusion of false helps summon (Christensen et al., 20).

[vi] Where keV equals kilo-electron volts, a unit of energy being observed.

[vii] See "Chandra Images and False Color" ed. Chandra EPO Group at SAO (Arcand, Watzke et al 2005) http://chandra.si.edu/photo/false_color.html

[viii] Images from http://sdo.gsfc.nasa.gov/data/ (November, 2011)

[ix] For more on this object see http://chandra.si.edu/photo/2006/bhcen/. Chandra EPO Group at SAO (Edmonds, Watzke, Arcand et al 2006)

[x] Examples of publicly-oriented open-FITS programs are at http://chandra.si.edu/photo/openFITS/, http://www.spacetelescope.org/projects/fits_liberator/datasets_archives/

[xi] http://astroart.cfa.harvard.edu/